# Near-energy-free Photonic Fourier Transformation for Convolution Operation Acceleration

Hangbo Yang[1,2,*], Nicola Peserico[1,2], Shurui Li[3], Xiaoxuan Ma[4], Russell L. T. Schwartz[1,2], Mostafa Hosseini[3], Aydin Babakhani[3], Chee Wei Wong[3], Puneet Gupta[3], Volker J. Sorger[1,2,*]

[1]Department of Electrical and Computer Engineering, University of Florida, Gainesville, FL, 32611, USA
[2]Florida Semiconductor Institute, University of Florida, Gainesville, FL, 32611, USA
[3]Department of Electrical and Computer Engineering, University of California Los Angeles, Los Angeles, CA, 90095, USA
[4]Department of Electrical and Computer Engineering, The George Washington University, Washington, DC 20052, USA

[*]Corresponding author. Email: hangbo.yang@ufl.edu, volker.sorger@ufl.edu

**Abstract:** Convolutional operations are computationally intensive in artificial intelligence services, and their overhead in electronic hardware limits machine learning scaling. Here, we introduce a photonic joint transform correlator (pJTC) using a near-energy-free on-chip Fourier transformation to accelerate convolution operations. The pJTC reduces computational complexity for both convolution and cross-correlation from $O(N^4)$ to $O(N^2)$, where $N^2$ is the input data size. Demonstrating functional Fourier transforms and convolution, this pJTC achieves 98.0% accuracy on an exemplary MNIST inference task. Furthermore, a wavelength-multiplexed pJTC architecture shows potential for high throughput and energy efficiency, reaching 305 TOPS/W and 40.2 TOPS/mm$^2$, based on currently available foundry processes. An efficient, compact, and low-latency convolution accelerator promises to advance next-generation AI capabilities across edge demands, high-performance computing, and cloud services.

**Main Text:** Convolutional neural networks (CNNs) have emerged as a cornerstone of modern artificial intelligence (AI) and machine learning, driving advancements in image recognition, natural language processing, and other data-intensive tasks (*1–5*). That is, the supermajority (>90%) of computational overhead (e.g. power) in CNNs is taken up by performing convolution operations (*6*, *7*); performing pattern recognition to raw data by extracting features.. At a higher abstraction level convolutions perform filtering of an input signal with a kernel to reveal relevant features and hence extract meaningful information from data. Unfortunately, convolution operations scale with a high computational complexity, **O**($N^4$), for two-dimensional ($N^2$) data (e.g. images) and same-size kernels, making them computationally expensive. Convolutions are routinely processed as time-serialized, kernel striding, processing of densified (not sparse) matrix-matrix multiplications which scale as **O**($N^3$). Optimizing this paradigm over the last decade has led to significant performance gains, i.e. 1000-fold, in tensor operation acceleration and was instrumental in unlocking the AI service revolution we are witnessing today (*8*). However, scalability limitations of this near brute-force approach are looming on the horizon, which have been recognized by data center and IT-related power consumption predictions (*9*), with a threat of heralding a possible AI 'recession', if not addressed by about 2030 (*10*).

To address these limitations and enabled continued AI services scalability, the convolution theorem can been considered to provide a solution; here a convolution operation is achieved as a straightforward dot-product multiplication, however, requiring to be inside the Fourier domain. The latter is costly for electronics as the transformation into- and out of the Fourier domain costs $\mathbf{O}(N^2\text{Log}(N))$ plus the dot product multiplication for two-dimensional data. In contrast, implementing the Fourier transformation (FT) in optics is straightforward using a lens. While free-space implementations of this concept, known as 4F systems, enables massive (~$10^6$ pixel) parallelism, it is limited to ~100 Hz slow update rates when spatial light modulators are used (*11*), and 10s' kHz using digital mirror display technology (*12–14*), such free-space systems do not offer the GHz fast update rates nor take advantage of the semiconductor scaling and manufacturing capabilities including advanced packaging.

Unlike, traditional computational filtering inside the Fourier domain (4F system) (*15, 16*), the complexity of a Fourier-based convolution operation ML accelerator can be further improved (i.e. reduced) by letting the kernel (i.e. filter) to cross-correlate with the data inside the Fourier domain. That is, the kernel enters 'front-end' alongside the data. This architecture, inspired by joint transform correlators (*17–19*), allows both the signal and kernel to reside in the same plane, eliminating the need for complex conjugation and relaxing alignment requirements (*16, 20*). Consequently, the design requires fewer components (e.g., it eliminates the need for complex-number data input and phase alignment components such as phase shifters at the first lens's back focal plane), simplifying implementation and maintenance, and potentially leading to lower fabrication costs and increased robustness when compared to a classical 4F system.

Here, we introduce an on-chip Fourier-based photonic joint transform correlator (pJTC) for efficient convolution operation acceleration. Novelties and innovations of the pJTC include, (i) rapid kernel and data programmability a million times faster (~GHz) compared to liquid crystal or micromirror-based approaches (~KHz); (ii) leveraging standard photonic integrated circuit components, trusted from optical transceiver technologies, with a novel process design kit (PDK) additions of an on-chip Silicon photonics-based Fourier lens (*21–23*); and (iii) chip-integrated lasers for computational parallelization via spectral multiplexing and compact design leveraging photonic wire bonding technologies (*24–26*). We experimentally test the fidelity of the photonic chip FT and address computational non-idealities via phase drift corrections and evaluate the machine learning capability of the pJTC in image classification task after training the system via an emulated-system model. Further, we show resilient operation by tuning the pJTC to account for hardware and data input/output- related non-idealities such as temporal jitter. We offer future system scale-out and scale-up performance opportunities towards large-scale implementation and computing pointing towards ~5x improvements over state-of-the-art compute-in-memory tensor-based systems, and ~100x over latest GPU's. This work demonstrates the feasibility of using pJTCs for ultra-fast and energy-efficient CNN processing. The pJTC's ability to perform convolutions with reduced computational complexity has the potential to herald a wide range of AI application acceleration from autonomous system to medical imaging (*27*).

Convolution, a fundamental operation in CNN, involves filtering an input signal with a kernel to extract features. This process, mathematically equivalent to cross-correlation in the context of CNNs, is computationally intensive for large datasets. It can be performed in either the spatial or Fourier domain (**Fig. 1A**). Spatial convolution (SC) is obtained by striding a kernel ($k^2$) across the signal matrix ($n^2$), a brute-force calculation with a computational complexity of $\mathbf{O}(N^2k^2)$. To improve efficiency, Fourier electrical convolution (FEC) based on Fast Fourier Transform (FFT) was developed, with a computational complexity of $O(N^2 \log N)$. Further enhancing efficiency, Fourier optical convolution (FOC) leverages the inherent parallelism of optics to perform Fourier transform and multiplication operations with high efficiency, achieving the lowest computational complexity of $O(N^2)$ (*12, 28, 29*). **Figure 1B** illustrates how FOC can be achieved by a joint transform correlator (JTC). In the JTC, the input signal $S(u,v)$ and kernel $K(u,v)$ are combined and passed through a lens to generate their combined Fourier transform in the optical domain, represented as: $E(u,v) = S(u,v) + K(u,v)$ [eq1]. A group of square-law detectors capture the intensity of this combined signal at the Fourier plane, resulting in: $I(u,v) = |S(u,v)|^2 + |K(u,v)|^2 + (|S(u,v)| \times e^{\varphi_s(u,v)}) \times (|K(u,v)| \times e^{-\varphi_k(u,v)+j(ud+vd)}) + (|S(u,v)| \times e^{-\varphi_s(u,v)}) \times (|K(u,v)| \times e^{\varphi_k(u,v)-j(ud+vd)})$ [eq2]. This intensity pattern $I(u,v)$ then passes through a second lens, producing the output as captured by a detector array. This output contains both the convolution of $S(u,v)$ and $K(u,v)$ (from the Fourier transform of the last two terms in eq2) and their auto-correlation (from the Fourier transform of the first two terms in eq2). Inspired by the optical JTC, we design a photonic JTC (pJTC), integrating all the necessary optical components - modulators, lenses, and photodetectors - on a single chiplet, to demonstrate an end-to-end feasible prototype of a pJTC accelerated convolution accelerator, which is a first steppingstone for the possibility for future volume production and system scalability.

The pJTC is an innovative approach that integrates photonic integrated circuit (PIC) components to enable Fourier-based data filtering on-the-fly. By leveraging light's properties, such as low latency (~ 10's ps), low energy consumption (passive FT), and parallelism (e.g. wavelength division multiplexing), the pJTC offers an elegant method to perform convolution operations for CNNs and analog signal processing, enabling rapid and energy-efficient chip implementations (**Fig. 2**). In this approach, the data for both the input and kernel are encoded using optical signals from components, here micro-disk modulators (MDMs), processed through the on-chip FT Fresnel-lenses, and ultimately detected using on-chip photodetectors (**Fig. 2A**). An off-chip 1550 nm continuous-wave (CW) laser is coupled into the pJTC via a grating coupler aligned to an attached fiber array. The light input is evenly split into 16 beams using four groups of beam splitters (BS). Each beam, representing either a signal element ($s_{1\sim8}$) or a kernel element ($k_{1\sim8}$), is modulated by one MDM. Modulation is achieved via a PN-junction, driven by an off-chip microcontroller or function generator. After modulation, the beams are guided by the waveguides and arrive at the front focal plane of an on-chip Fresnel-lens based on a silicon-on-insulator platform. The combined signal and kernel beams interfere as they propagate through the first Fresnel-lens, undergoing a Fourier transform at the back focal plane. The optical signal is then sampled by 16 waveguides and detected by 16 photodetectors (PDs), implementing a nonlinear transformation (square function) in the electrical domain. The resulting electrical signals are collected by the microcontroller and used to drive another set of 16 MDMs, modulating a second set of optical beams. These beams propagate through a second on-chip Fresnel-lens and are detected by a second group of 16 PDs, completing the inverse Fourier transform and yielding the convolution of the signal and kernel. Based on the design above, the pJTC chiplet was fabricated through AIM Photonics, showcasing the integrated photonic components and waveguide routing (**Fig. 2C**). A custom printed circuit board (PCB) was developed to interface the pJTC chiplet with its electrical controllers (heaters, monitoring photodetectors), low-speed, low-energy transimpedance amplifiers (TIAs), digital-to-analog converters (DACs), and analog-to-digital converters (ADCs). This PCB enables integration of the pJTC chiplet into the electronic data system, facilitating control, readout, and connection to other components. The packaged pJTC, shown with an 8-channel fiber array connected to the PCB (**Fig. 2B**), provides a compact and integrated solution for optical signal processing.

Following fabrication by AIM Photonics, the pJTC chiplet we operationally tested the optoelectronic components and the entire prototype system for convolution functionality. First, initial calibration involved aligning the resonance of each MDM with the common input laser wavelength by precisely tuning the integrated heaters. This iterative process accounted for thermal crosstalk between closely spaced heaters. One measured MDM spectrum exhibited a Q-factor of 4500 and an extinction ratio (ER) of ~10 dB (**Fig. 2D**), confirming its suitability for modulation and providing sufficient isolation to minimize wavelength-crosstalk and maximize the signal-to-noise ratio (SNR). Second, to verify the PN-junction modulation functionality of the MDMs, the measured power difference between the output and input was compared with simulations performed in Ansys-Lumerical Interconnect (**Fig. 2E**). Noise measurements were conducted at the receiver side, varying the input modulator intensity to determine the ER and noise level of each PD. Eye diagrams in the insert of **Fig. 2E** at 1 GHz operation showed an open eye indicative of sufficient signal integrity and low jitter at those speeds and modulator format. The low power difference, attributed to the low input laser power required for increased power efficiency, aligned well with simulation results, confirming the efficient operation of the MDMs. Further calibration of modulation intensity of each MDM ensured multi-level modulation capability across all MDMs. The performance of the pJTC's on-chip Fresnel-lens was simulated using Ansys-Lumerical FDTD (see supplementary information). FTDT simulation results (**Fig. 2F,G**) demonstrate efficient light focusing and directing, while also revealing diffraction effects from the input waveguides. To mitigate crosstalk arising from boundary reflections, the Fresnel-lens width was optimized to balance footprint and crosstalk suppression. The results validate the effectiveness of the on-chip Fresnel-lens in performing Fourier transformations, a crucial aspect of the pJTC's convolution operation.

The aforementioned component-level measurements enable the determination of bit resolution for each receiver; to assess the stability of each input in the pJTC, we conducted 30 repeated measurements of the output differential of photodetectors (PDs) by cycling between all-0 inputs and all-1 inputs, capturing the results in both states (**Fig. 2H**). The obtained standard deviation of 78.9 nW is less than 4% of the average output differential, demonstrating decent stability. Additionally, a stability analysis of the combined voltage signal from all PDs over a 24-hour period shows in **Fig. 2I**. Consistent performance was observed with constant input voltages applied to the MDMs: $V_{min}$ = -1.5 V for $Input_0$ (all-0 state) and $V_{max}$ = -0.2 V for $Input_1$ (all-1 state). The stable ER throughout this period points to long-term operational stability of the pJTC chiplet and packaged approach. To evaluate stability across a wider range of input values, all possible input combinations across the 16 MDMs ($2^{16}$ = 65536) were tested (**Fig. 2J**), each modulated with a 1-bit signal. The aggregated voltage output from the PDs exhibited 16 distinct periods, confirming the pJTC's reliability and consistent operation.

Overall, initial experimental and simulation results at both the component and system levels demonstrate the successful fabrication, integration, and operation of the pJTC chiplet, with observed performance characteristics, including efficient modulation, low crosstalk, accurate Fourier transformation, and stable operation, thus paving the way for the pJTC's applications in an controlled environment (as tested). Beyond stable data-center environments, future tests of emerging photonic ML accelerators such as the pJTC, shall include impacts from thermal and other harsh-environments, especially for edge applications (e.g. thermal shock, humidity, rad-hard etc).

The on-chip Fresnel-lens is a key component of the photonic JTC system, and its functional accuracy directly impacts overall performance in terms of convolution accuracy (**Fig. 3**). Experimental characterization of the lens (yellow line in **Fig. 3B** and third column of the MNIST handwritten digits in **Fig. 3C**) reveal a deviation from the ideal Fourier transform (blue line in **Fig. 3B** and second column of the MNIST handwritten digits in **Fig. 3C**). This requires a one-time calibration step, rationalized and performed as follows; this discrepancy can be attributed to various factors, including lithographic limitations in lens fabrication and variations in the silicon substrate (thickness, roughness) across the Fresnel-lens footprint. We focused on mitigating the impact of the optical effective path length differences between input waveguides, which can arise even after careful waveguide length equalization due to inherent silicon variations from the foundry process. To address this, we developed a model (**Fig. 3A**) that incorporates the amplitude and phase contributions from the MDMs, a phase term accounting for path differences, and the Fourier transform performed by the Fresnel-lens. The modulator parameters were obtained from the foundry's process design kit (PDK), while the Fresnel-lens was modeled as an ideal Fourier transform. To determine the phase shifts caused by path differences, we collected a comprehensive dataset of input-output combinations and employed an optimization algorithm to minimize the error between the model and experimental results by adjusting the phase terms. The pink line in **Fig. 3B** demonstrates the effectiveness of this approach, showing a close match between the experimental outcome after phase correction and the ideal Fourier transform model (blue line in **Fig. 3B**). Furthermore, **Fig. 3C** illustrates the impact of this correction on the recognition of MNIST handwritten digits. While the "actual" performance is initially suboptimal, it can be effectively calibrated. This calibration process accounts for the optical path length differences in each channel, with a computational complexity dependent on the number of channels and the bit resolution per channel. Once determined, these static optical path length differences are compensated for using photonic phase shifters. The results obtained after phase correction closely resemble those of an ideal FT system, confirming that accurate CNN operation is achievable despite the non-ideal optical effective path differences. For future implementations, in addition to precise waveguide balancing, we envision incorporating phase-change material (PCM)-based components (*30, 31*) as energy-efficient on-chip calibration channels, enabling non-volatile tuning of individual optical paths, offering further refinement and control over the system's performance.

Next, we investigate the obtainable machine learning training fidelity by pJTC chiplet emulation (**Fig. 4A**); we trained our neural network off-chip using a comprehensive physical model that accurately captures the electro-optical system's behavior, including non-idealities such as waveguide sampling effects, electrical signal delays, MDM modulating depth, and SNR of the PDs (see supplementary material, Section 1). After validating the model through simulations and retraining the fully connected (FC) layer to compensate for any discrepancies, we achieved high classification accuracy on the MNIST dataset. Due to electrical path length deviations in our PCB, each channel experiences temporal delays in the input electrical signals. We estimate these deviations to be approximately 10% for RF signals at 10 GHz and 1% for RF signals at 1 GHz. To assess the impact of these delays, we emulated the system with four different levels of random temporal delay: no, 1%, 5%, and 10%. For the MNIST dataset, we observed an accuracy of 98.0% with no temporal delay and 97.2% with 1% random temporal delay, which monotonically declined to 95.3% with 10% random temporal delay. The minimal accuracy drops of only 1.9% as the delay increases from 1% to 10% highlights the robustness of our system to electrical signal delays (**Figs. 4B–4E**).

Automation and intelligent computing have become the single most dominant driver of the digital age in 21st century. Specifically, Artificial Intelligence (AI) and Machine Learning (ML) business products have become an economic force; 80% of all corporations reported having at least one AI product, trend rapidly rising (*32*). This parallels the demand for data center units (DCU) expansions with another 80 GW projected to be installed by 2030 in the USA (*32*). Such compute and AI demand requires novel devices, chiplet, package, supply chain, and ecosystem approaches. Notably, much of the 1000× compute performance gains in GPU's over the past decade were based on other than FET scaling improvements (*8*). To unlock the next 1000× AI and HPC gains, innovations in chip packaging, architecture, 3D heterogeneous integration of electronic & photonic chips alike, a leaner and more secured supply chain, are needed.

The rapid advancement of high-performance computing (HPC) and AI applications, particularly in ML and deep learning (DL), has led to an exponential increase in the demand for high-bandwidth, low-latency, and energy-efficient communication between processing cores on a chip. Traditional electrical interconnects are struggling to keep up with these demands due to their limited bandwidth, high latency, and significant power consumption riven from charging and discharging densely coupled and capacitance-generating wirings.

To provide a forward looking benchmark of a photonic passive Fourier-based convolution accelerator against state-of-the-art ML systems **Figure 5** is an attempt; operating a single pJTC chiplet prototype at 10 GHz gives a computational footprint density of 0.16 TOPS/mm² and a computational efficiency of 3.0 TOPS/W, which is below electronic counterparts like NVIDIA GPU A100, H100 and B200 (*33*) as well as academic convolutional accelerators (*34*, *35*). To unlocking the full potential of a scaled pJTC, we optimized the on-chip Fresnel-lens by adjusting its input waveguide pitch and slot dimensions (**Figures S1-S3**). To further enhance parallel processing capabilities, we propose an upgraded pJTC architecture incorporating wavelength division multiplexing (WDM), as shown in **Fig. S4**. This design enables computational parallelization through chip-integrated lasers utilizing spectral multiplexing and a compact layout leveraging photonic wire bonding technologies (*24–26*). The system features $n$-wavelength and $m$-channel input signals and kernels modulated by MDMs, optimizing efficiency and scalability. The same signal is shared among all the $n$-wavelength to save the number of MDMs to reduce the power consumption while maintaining the same throughput. Each waveguide carries $n$-wavelength optical signals, which are then separated by additional $n * m$ MDMs before being collected by $n * m$ PDs. The electrical outputs from the PDs are summed channel-by-channel across all groups. These summed results then modulate MDMs at a specific wavelength, pass through another on-chip Fresnel-lens, and are finally detected by a group of PDs. **Figure 5** presents the performance comparison of the upgraded pJTC with state-of-the-art electronic neural networks (*33–35*). Computational efficiency improves with increasing channel number ($m$), but computational footprint density begins to decrease beyond 16 channels (gray dash lines in **Fig. 5**, details in **Fig. S5**). This decrease is attributed to the footprint increasing faster than the absolute computational speed, particularly due to the Fresnel lens's footprint (pie charts in **Fig. 5**). In contrast, both computational efficiency and computational footprint density increase with wavelength number ($n$). This is because the Fresnel-lens's footprint remains constant, and the footprint increase of other components is slower than the gain in absolute computational throughput. Specifically, many MDMs and DACs for different wavelengths at the input planes of both on-chip Fresnel-lenses can be shared. Considering SOTA PDK on-chip electrical components, one can achieve: 305 TOPS/W accelerator (256 Ch#, 32 wavelengths) and 40.2 TOPS/mm² accelerator (32 Ch#, 32 wavelengths). Excluding electrical components, i.e. only the photonic chiplet offers 987 TOPS/W accelerator (256 Ch#, 32 wavelengths) and 50.8 TOPS/mm² accelerator (32 Ch#, 32 wavelengths). For detailed assumptions and calculations see Section V of the supplementary online material. To increase the computational footprint density of the pJTC with a higher channel number, optimizing the Fresnel-lens is crucial, which can be realized by: (1) boundary absorption techniques, leveraging principles from stealth RF technology to minimize 1D lens width, potentially through absorption shape/metamaterial design for wide-angle performance; (2) lens folding in length, especially when the propagation width is significantly less than the length. To improve computational energy efficiency of the pJTC further, minimizing optical-to-electrical (O-E) and electrical-to-optical (E-O) conversion losses is crucial. Advancements in pJTC technology hold the potential to revolutionize real-world applications that rely on CNNs, including autonomous driving, medical imaging, and natural language processing. The ultra-fast and energy-efficient processing capabilities of pJTCs can enable these applications to tackle more complex tasks with greater speed and accuracy.

## Figures

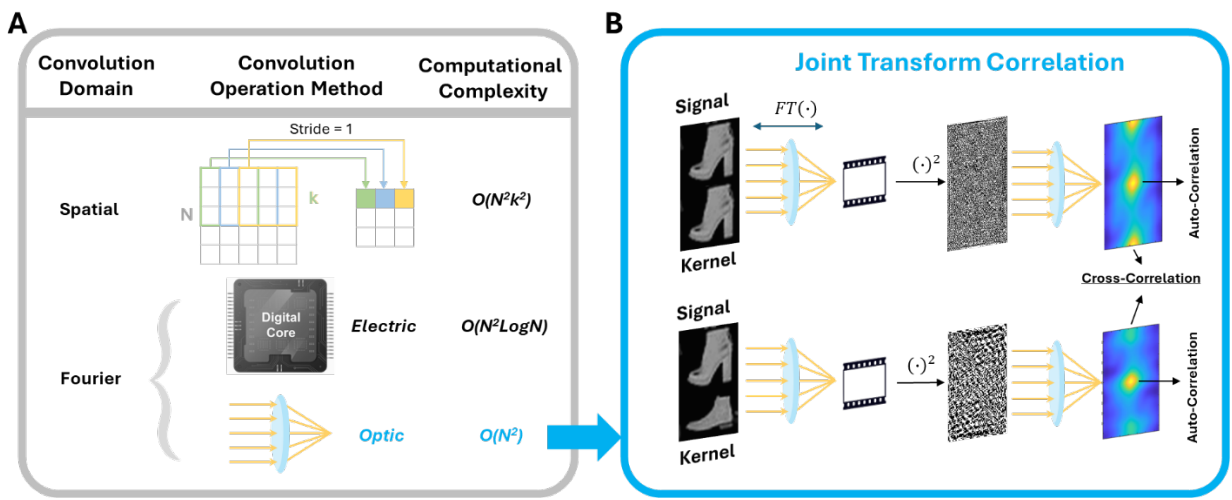

**Fig. 1. Concepts to perform convolution operations forming the basis for many machine learning algorithms**. (**A**) Comparison of spatial convolution (SC), Fourier electrical convolution (FEC), and Fourier optical convolution (FOC) in terms of computational complexity. $N$ data input, $k$ kernel size. Exemplary, shown a two-dimensional input, $N^2$ (e.g. image). Reduced complexity scaling using an optical chip can be achieved via FOC and photonic foundry processes enable chip-based implementation options, as demonstrated in this work. (**B**) FOC can be utilized in a joint transform correlator (JTC) by optically generating the Fourier transform of the combined input Signal and Kernel, detecting the intensity pattern, and producing the auto- and cross-correlation between Signal and Kernel. Unlike a classical 4F optical system, the JTC offers fewer components, simplifying implementation and system stability, and reduced fabrication complexity.

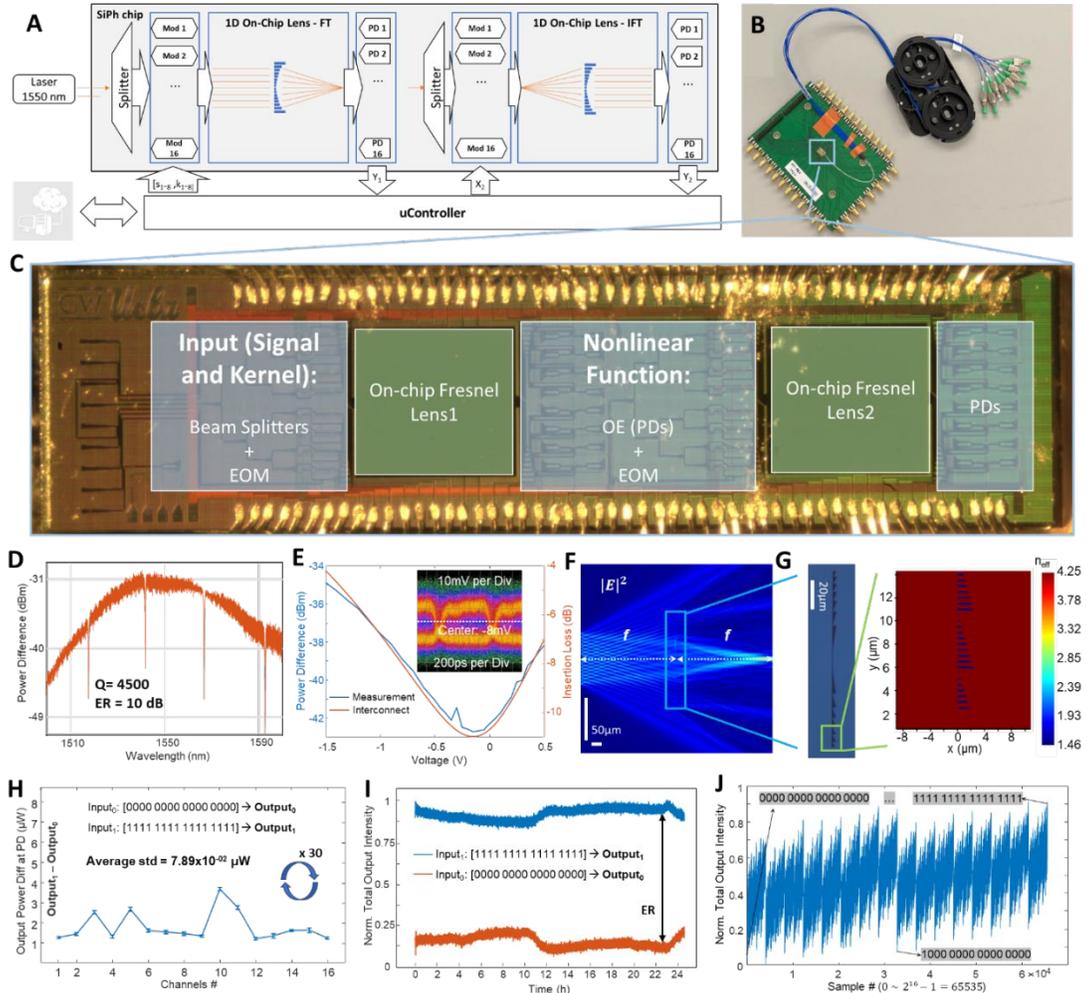

**Fig. 2. Design, Package and Test of the photonic Joint Transform Correlator (pJTC) Chiplet and Prototype.** (**A**) Schematic of the pJTC including silicon photonic (SiPh) chiplet and off-chip controller, (**B**) Packaged pJTC with the SiPh chiplet, custom PCB and an 8-channel fiber array. (**C**) Optical microscope image of the fabricated SiPh chiplet from AIM Photonics. (**D**) The measured spectrum of the pJTC MDM exhibits a Q-factor of 4500 and an extinction ratio of ~10 dB, demonstrating its suitability for modulation and providing sufficient isolation to minimize wavelength-crosstalk. (**E**) Comparison of the measured power difference between the MDM output and input with the simulated response in Ansys-Lumerical Interconnect as a function of the bias voltage on MDM. Insert: Eye diagram of the pJTC PD with MDM modulated at an operating speed of 1 GHz, illustrating clear signal integrity and low jitter. (**F**) Simulation results of the pJTC on-chip Fresnel-lens obtained using Ansys-Lumerical FDTD, showing efficient light focusing and directing. (**G**) Optical microscope image of the on-chip Fresnel-lens and refractive index of its part. (**H**) 30 repeated measurements of the output differential of photodetectors (PDs) by cycling between all-0 inputs and all-1 inputs, yielded a standard deviation of 78.9 nW, demonstrating high stability in performance. (**I**) Stability analysis of the combined voltage signal from all the PDs over a 24-hour period, demonstrating consistent total ER while cycling between $Input_0$ (all-0 state) and $Input_1$ (all-1 state). (**J**) Aggregated voltage output from the PDs for all possible input combinations across 16 MDMs ($2^{16}$ = 65536), each modulated with a 1-bit signal. The 16 distinct periodic signals observed confirm the reliability and consistent operation of the pJTC.

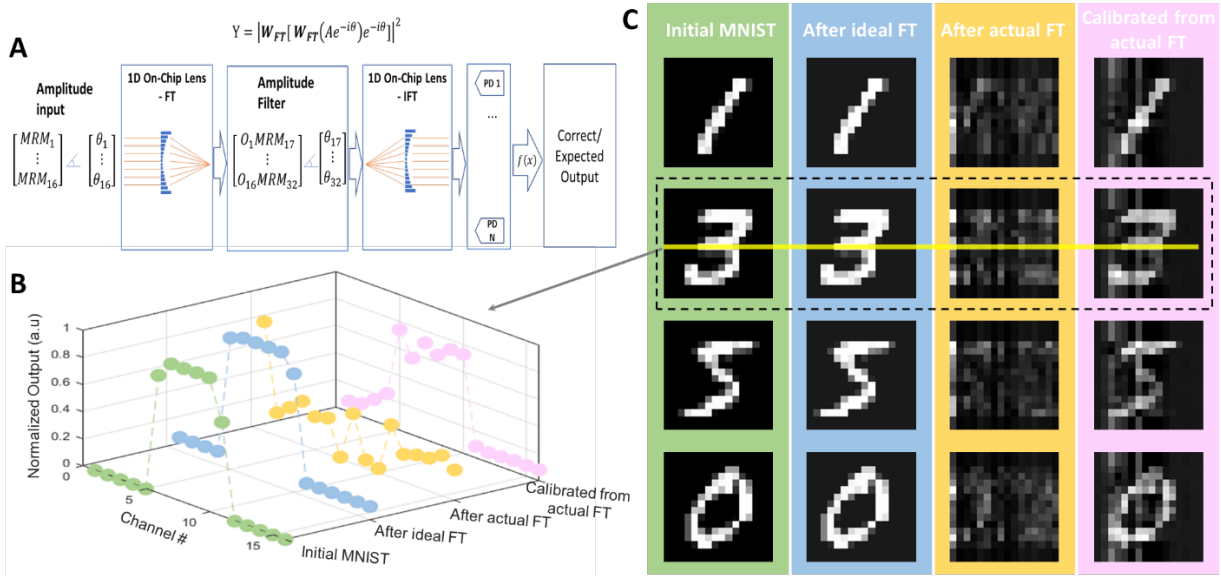

**Fig. 3. Phase correction for the on-chip Fresnel-lens.** (**A**) Model incorporating sub-micron path differences between input waveguides before the Fresnel-lens enabling stabilized image classification after an initial calibration step as follows (see supplementary material for further details): (**B**) Comparison of a representative cross-section taken from the same MNIST digit image ("3" shown in panel (**C**) at various stages of processing: the initial MNIST image (green line), the output after an ideal Fourier transform (blue line), the output after the actual on-chip lens Fourier transform (yellow line), and the calibrated output obtained from the actual on-chip lens after applying phase correction (pink line).

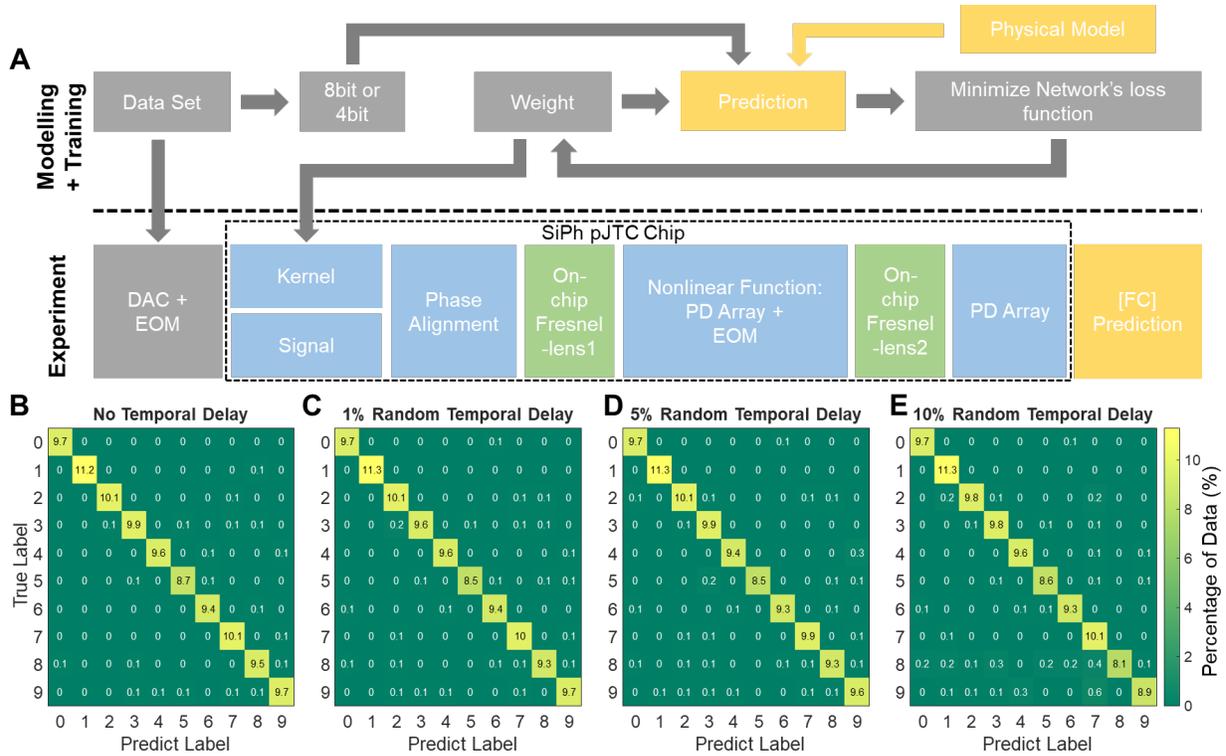

**Fig. 4. JTC-based Convolutional Neural Network (CNN) Inference Test and Temporal (jitter) Signal Analysis**. (**A**) A physical model of the input-kernel layer is used to train the entire system and obtain the optimal weights for the kernel, which are then loaded into the input-kernel channels. Experimentally obtained results from the input-kernel layer are fed to the fully connected (FC) layer for performing the final prediction on unseen data. (**B-E**) Confusion matrices as a junction of channel-to-channel jitter from emulation demonstrate high degree of stability of the pJTC on a mock dataset (MNIST) classifying handwritten digits. The matrices show the classification accuracy for 10,000 test images with zero, 1%, 5%, and 10% random delay introduced in the input electrical signal, achieving total accuracies of 98.0%, 97.2%, 96.9%, and 95.3%, respectively.

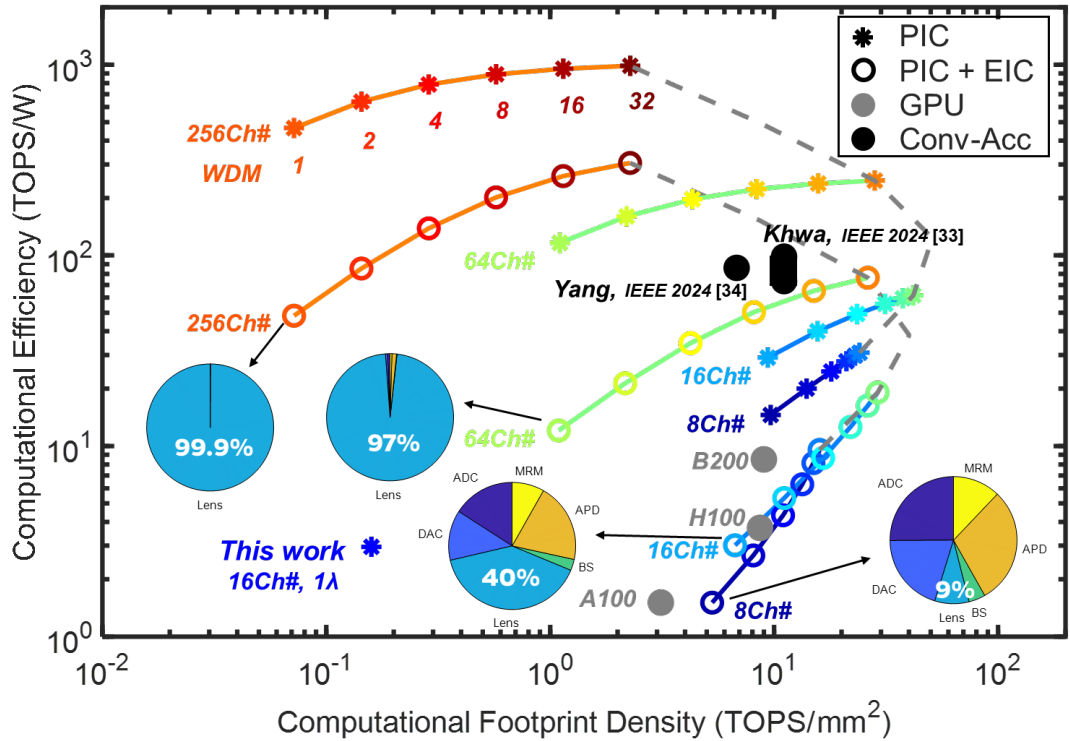

**Fig. 5. Machine Learning Accelerator Comparison**. Computational efficiency (TOPS/W) and footprint density (TOPS/mm²) of a wavelength-division multiplexed (WDM)-parallelized pJTC architectures vs. electronic neural GPUs (gray dots) and academic convolutional accelerators (black dots), demonstrating the effects of varying channel (*Ch#*) and wavelength (*WDM*) numbers. Insets show a growing Fresnel-lens footprint contribution on the chip area, making the case for mid-size photonic pJTC chiplets for optimized throughput-area efficiency. Dashed lines highlight the decrease in computational footprint density beyond 16 channels. This comparison also details achieved performance with solid (star-shape) vs. open (circle-shape) dots for including the electrical drive and mixed-signal conversion circuitry. For details see supplementary information, Section V.